\documentclass[prd,twocolumn,showpacs,floatfix]{revtex4}
\usepackage{hyperref,amssymb,amsmath,mathrsfs,bm,graphicx}
\usepackage{enumitem}
\usepackage{color}
\begin{document}
\title{General relativistic polytropes for anisotropic matter:\\ The general formalism and applications}
\author{L. Herrera}
\email{lherrera@usal.es}
\altaffiliation{Also at U.C.V., Caracas}
\affiliation{Departamento de F\'\i sica Te\' orica e Historia de la Ciencia,
Universidad del Pa\'\i s Vasco, Bilbao 48940, Spain}
\author{W. Barreto}
\email{wbarreto@ula.ve}
\affiliation{Centro de F\'\i sica Fundamental, Facultad de Ciencias,
Universidad de Los Andes, M\'erida 5101, Venezuela}
\begin{abstract}
We set up in detail the general formalism to model polytropic general relativistic stars with anisotropic pressure.  We shall consider  two different possible polytropic equations, all of which yield the same  Lane-Emden equation in the Newtonian limit.  A heuristic model based on an ansatz to obtain anisotropic matter solutions  from known solutions for isotropic matter is adopted to illustrate the effects of the pressure anisotropy on the structure of the star.  In this context,  the Tolman mass, which is  a measure of the active gravitational mass, is invoked to explain some features of the models.  Prospective  extensions  of the proposed approach are pointed out.
\end{abstract}
\date{\today}
\maketitle

\section{Introduction}
The polytropic equations of state have a long and a venerable history,  they were introduced in the context of Newtonian gravity, in order to deal with a variety of astrophysical problems (see Refs. \cite{1, 2, 3, 9, 8,10} and references therein),  their great success stemming mainly  from  the simplicity of the equation of state and the ensuing main equation (Lane-Emden).  

The theory of polytropes is based on the polytropic equation of state, which in the Newtonian case reads
\begin{equation}P=K\rho_0^{\gamma}=K\rho_0^{1+1/n} ,\label{Pol}\end{equation}
where $P$ and $\rho_0$ denote the isotropic pressure and the  mass (baryonic) density,  respectively. Constants $K$, $\gamma$, and $n$ are usual called  the polytropic constant, polytropic exponent, and polytropic index, respectively.

The polytropic equation of state may be used to model two very different types of situations, namely:
\begin{enumerate}[label=(\roman*)]
\item When the polytropic constant $K$ is fixed and can be calculated from natural constants. This is the case of a completely degenerate gas in the nonrelativistic ($\gamma=5/3; n=3/2$) and relativistic limit ($\gamma=4/3; n=3$). Polytropes of this kind are particularly useful to model compact objects such as white dwarfs (WDs), and they lead in a rather simple way to the 
Chandrasekhar mass limit.
\item When $K$ is a free parameter, as, for example, in the case of isothermal ideal gas, or in a completely convective star.
Models related to isothermal ideal gas are relevant in the so-called Sch\"onberg--Chandrasekhar limit (see Ref. \cite{3} for details).
\end{enumerate}

However in some cases the object under study may be  compact enough as to require the use of general relativity (e.g. neutron stars). In the context of this theory, polytropic equations of state have also been widely used (see Refs. \cite{4a,4b,4c,5,6,7,11,12,13}  and references therein). 

Both in the Newtonian and in the general--relativistic regime, the fluid under consideration is generally assumed  to be endowed with isotropic pressure (Pascal principle). However in a recent work \cite{HB} we have extended the framework  used to deal with Newtonian polytropic equations of state to the case of anisotropic fluids (principal stresses unequal).

The motivation to undertake such an  endeavour was  based on the fact that the local anisotropy of pressure may be caused by a large variety of physical phenomena of the kind  we expect to find  in compact objects (see Ref. \cite{14, hdmost04, hls89, hmo02, hod08, hsw08, p1, p2, p3} (and references therein  for an extensive discussion on this point). 

Among all possible sources of anisotropy,  there are two  particularly related to our primary interest. The first one is the intense magnetic field observed in compact objects such as white dwarfs,  neutron stars, or magnetized strange quark stars  (see, for example, Refs. \cite{15, 16, 17, 18, 19} and references therein). Indeed, it is a well-established fact  that  a magnetic field acting on a Fermi gas produces pressure anisotropy (see  Refs. \cite{ 23, 24, 25, 26} and references therein). In some way, the magnetic field can be addressed as a fluid anisotropy. 

 Another source of anisotropy expected to be present  in neutron stars and, in general, in highly dense matter,  is the viscosity (see \cite{Anderson, sad, alford, blaschke, drago, jones, vandalen, Dong} and references therein).

An alternative approach to anisotropy comes from kinetic theory
using the spherically symmetric Einstein-Vlasov equations, which admits a very rich class of static solutions, none of them isotropic (Ref. \cite{ref1, andreasson, ref2} and
references therein). The advantages or disadvantages of either approach are related to the specific problem under consideration. 

{Based in  all the arguments above, it is our main purpose  in this paper to develop the general formalism to describe polytropes in the presence of pressure anisotropy, within the context of general relativity. }

For the  sake of completeness, we shall first  review very briefly the theory of polytropes for a perfect (isotropic) general relativistic fluid. In this latter case  there exist two possible versions of the polytropic equation of state leading to the same Newtonian limit.

 Next, we shall display the general formalism  for anisotropic fluids. In this case,  we shall consider the same two possible polytropic equations of state of the isotropic case, but now applied to the radial pressure alone. Both cases share the same Newtonian limit.

{ In  order to integrate the obtained system of equations we need to provide further information about the anisotropy inherent to the problem under consideration. For doing that, we shall  assume an ansatz,  allowing us to specific modeling. For these models we shall also calculate the Tolman mass, whose behaviour allows one to understand some of their peculiar features. As we shall see below, our method  links our models continually with the isotropic case (see Sec. IV),
thereby allowing us to bring out the influence of anisotropy on the structure
of the object. However it should be stressed that the above mentioned models are  presented with the sole  purpose to illustrate the method, the natural way to obtain models, which consists in providing the specific information about the kind of anisotropy present in each specific problem.}
Finally, we shall conclude with a summary of results and some  possible extensions of our formalism.

\begin{figure}
\includegraphics[width=3.in,height=4.in,angle=0]{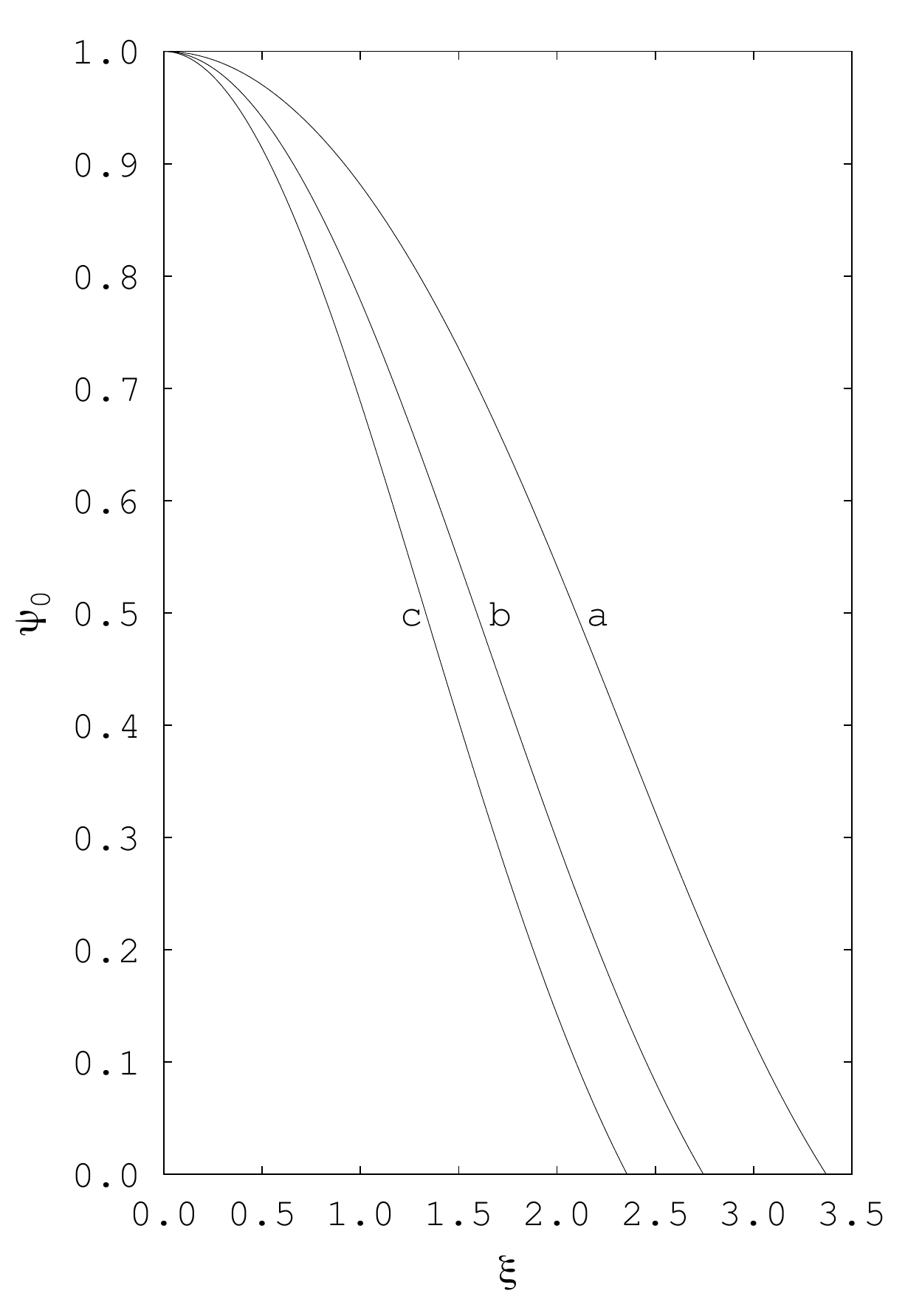}
\caption{$\psi_0$ as a function of $\xi$ for $\alpha=0.1$, $n=1$ and $h=0.5$ (curve a); $h=1.0$ (curve b); $h=1.5$ (curve c). For a wide range of $(\alpha,n,h)$ this figure is qualitatively representative.}
\label{fig:I2}
\end{figure}
\begin{figure}
\includegraphics[width=3.in,height=4.in,angle=0]{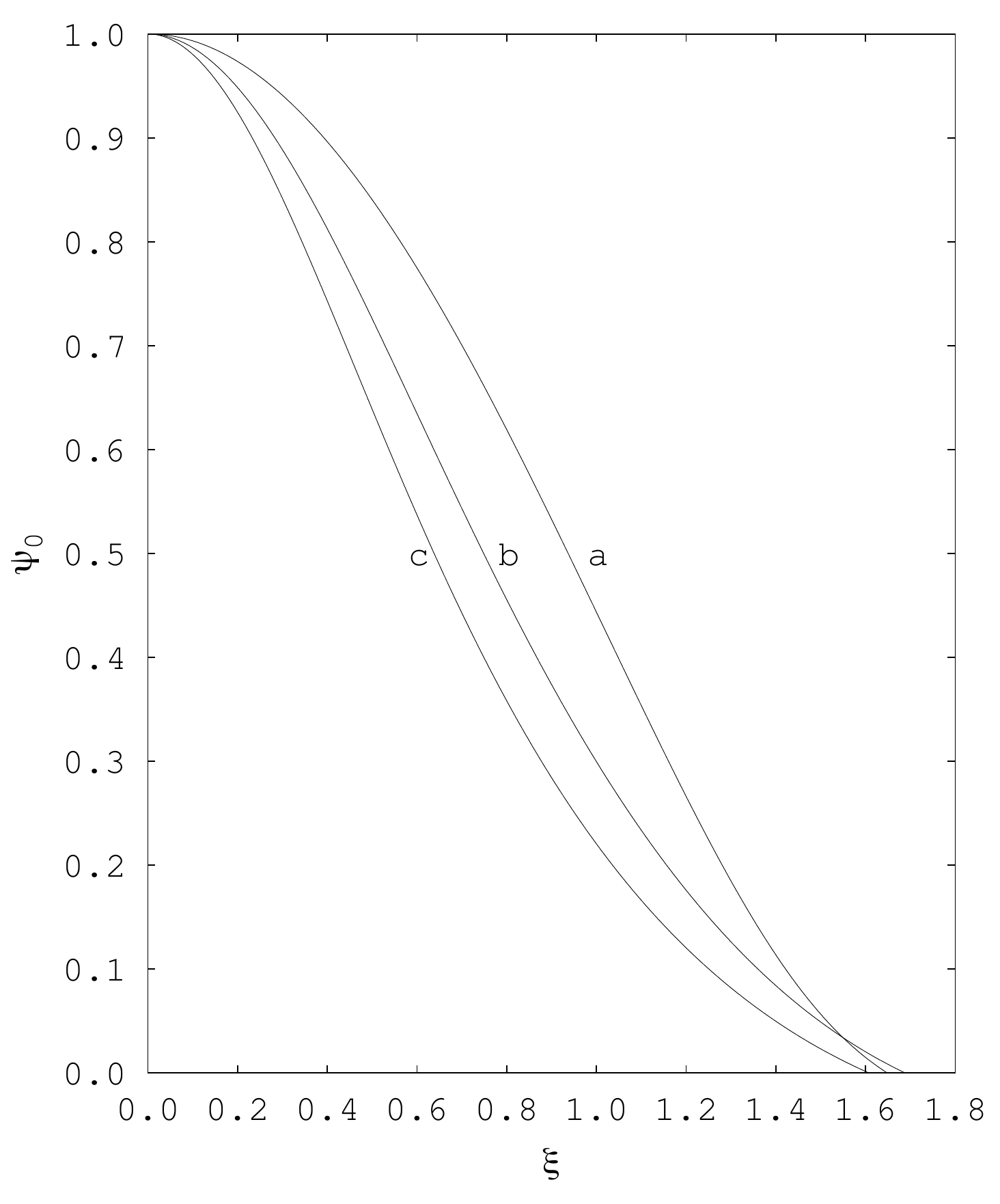}
\caption{$\psi_0$ as a function of $\xi$ for $\alpha=1$, $n=0.5$ and $h=0.5$ (curve a); $h=1.0$ (curve b); $h=1.5$ (curve c). Note the anomaly for $h=0.5$, that is, $\psi_0(\xi;h=1) > \psi_0(\xi;h=0.5)$ in some sector close to the surface $\xi_\Sigma$.}
\label{fig:I10}
\end{figure}
\begin{figure}
\includegraphics[width=3.in,height=4.in,angle=0]{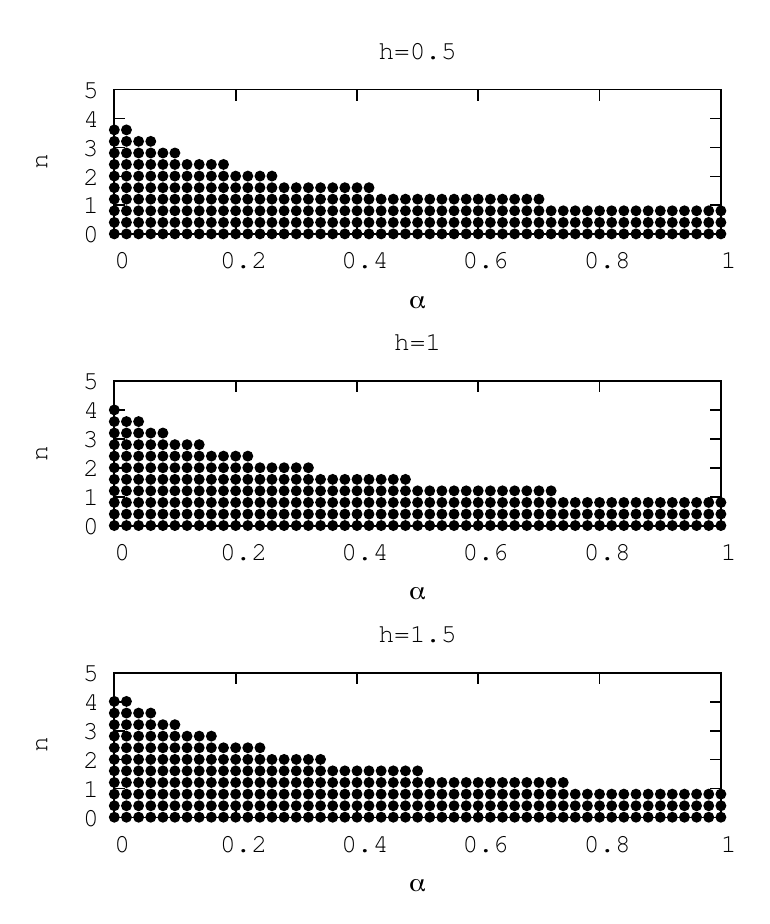}
\caption{Case I: Duplets $(\alpha,n)$ for bounded sources with $h$: 0.5$; 1.0$; $1.5$. Clearly $h>1$ favors the bounded configurations whereas  $h<1$ does not}
\label{fig:I6}
\end{figure}
\begin{figure}
\includegraphics[width=3.in,height=4.in,angle=0]{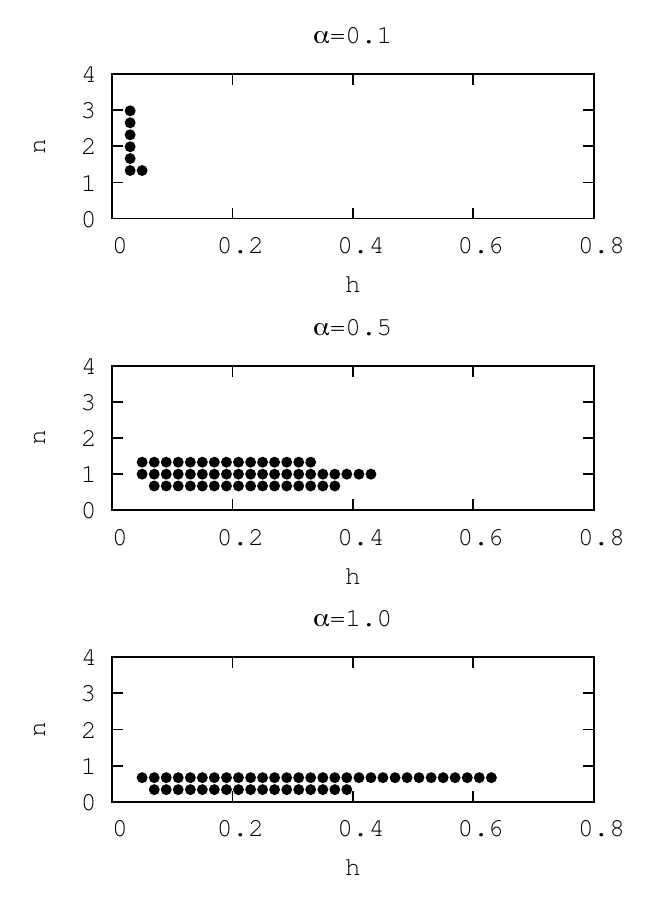}
\caption{Case I: Duplets $(h,n)$ for different $\alpha$ of bounded sources contained in figure \ref{fig:I6}, which show anomalous behaviour like in figure \ref{fig:I10}.  To find this subset of anomalous models, we do a search of parameters for which the behaviour appears at least in some interior point of the distribution.}
\label{fig:Ianoma}
\end{figure}
\begin{figure}
\includegraphics[width=3.in,height=4.in,angle=0]{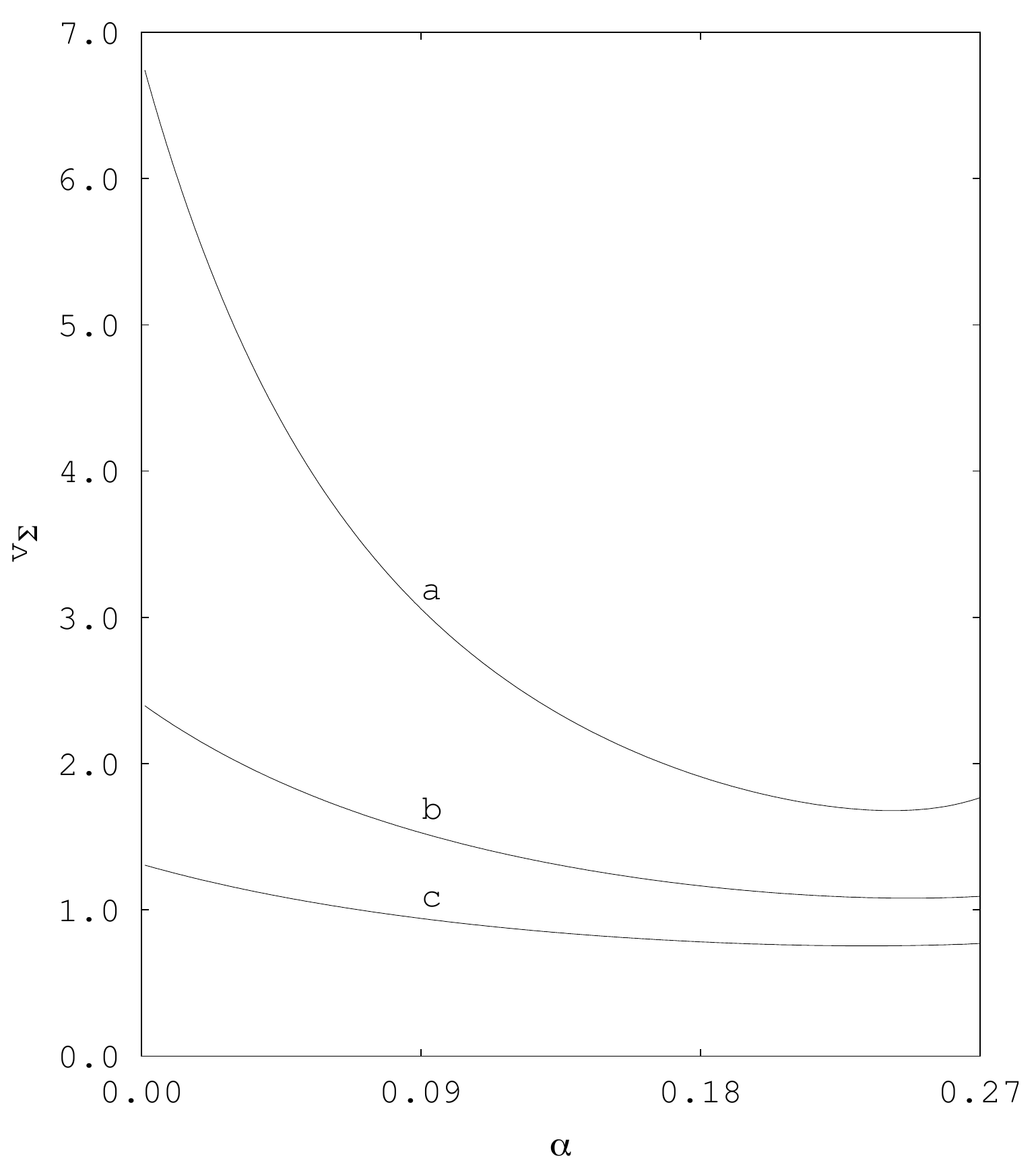}
\caption{Case I: $v_\Sigma$ as a function of $\alpha$ for $n=2$ and $h=0.5$ (curve a); $h=1.0$ (curve b); $h=1.5$ (curve c). In this case the limit value for $\alpha$ is $0.27$ to get a bounded source. For a wide range of $(n,h)$ this figure is qualitatively representative.}
\label{fig:I4}
\end{figure}
\begin{figure}
\includegraphics[width=3.in,height=4.in,angle=0]{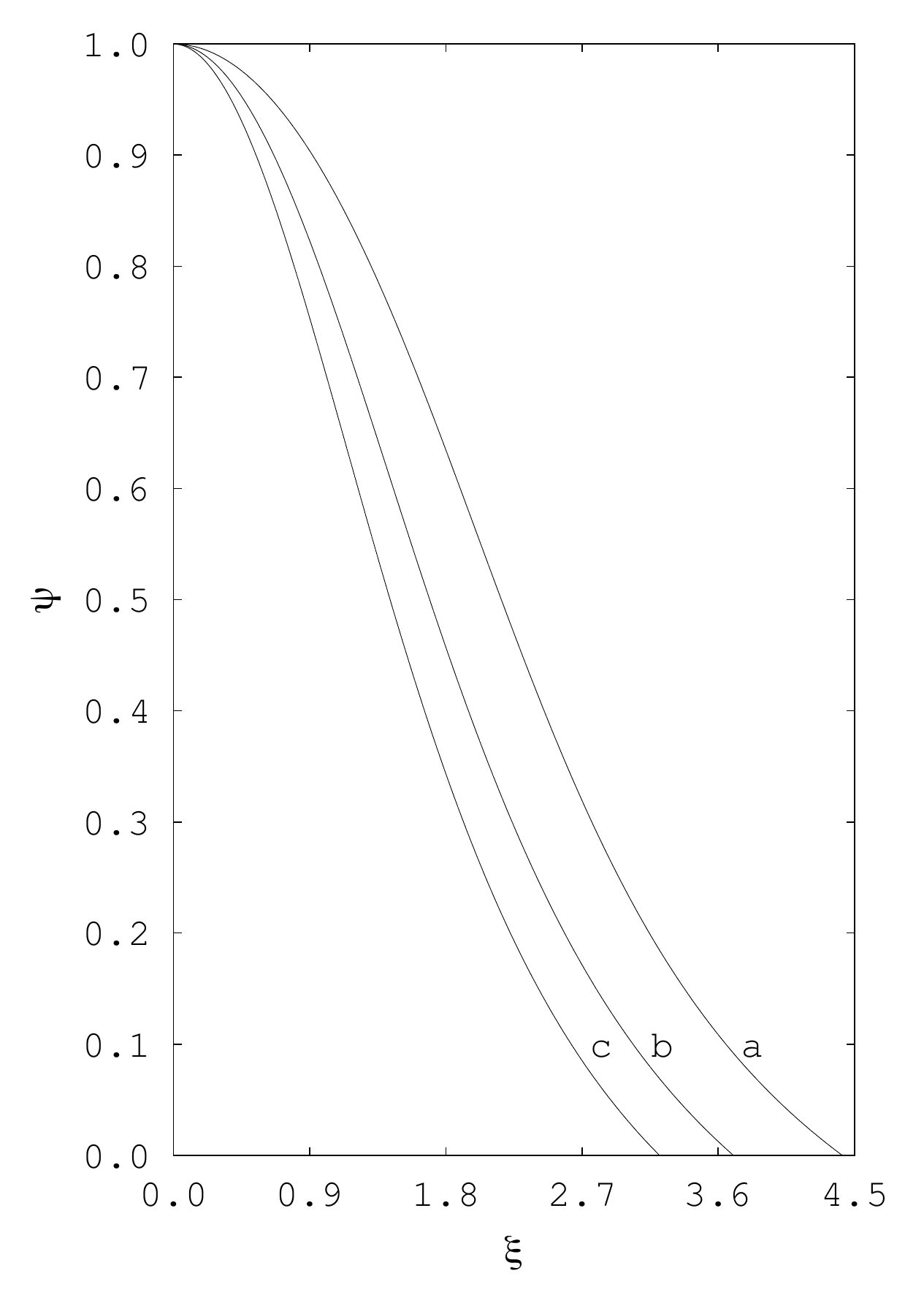}
\caption{$\psi$ as a function of $\xi$ for $\alpha=0.1$, $n=2$ and $h=0.5$ (curve a); $h=1.0$ (curve b); $h=1.5$ (curve c). For a wide range of $(\alpha,n,h)$ this figure is qualitatively representative.}
\label{fig:II1}
\end{figure}

\section{The general relativistic polytrope for a perfect fluid}\subsection{The field equations}
We consider spherically symmetric static distributions of fluid, bounded by a spherical surface $\Sigma$. \noindent
 The line element is given in Schwarzschild--like coordinates by 
\begin{equation}ds^2=e^{\nu} dt^2 - e^{\lambda} dr^2 -r^2 \left( d\theta^2 + \sin^2\theta d\phi^2 \right),\label{metric}\end{equation}\noindent 
where $\nu$ and $\lambda$ are functions of $r$.
We number the coordinates: $x^0=t; \, x^1=r; \, x^2=\theta; \, x^3=\phi$. We use geometric units and therefore we have $c=G=1$.
 The metric (\ref{metric}) has to satisfy Einstein field equations which in our case read \cite{hdmost04}:
\begin{equation}\rho=-\frac{1}{8 \pi}\left[-\frac{1}{r^2}+e^{-\lambda}\left(\frac{1}{r^2}-\frac{\lambda'}{r} \right)\right],\label{fieq00}\end{equation}
\begin{equation}P =-\frac{1}{8\pi}\left[\frac{1}{r^2} - e^{-\lambda}\left(\frac{1}{r^2}+\frac{\nu'}{r}\right)\right],\label{fieq11}\end{equation}
\begin{eqnarray}P = \frac{1}{8 \pi}\left[ \frac{e^{-\lambda}}{4}\left(2\nu''+\nu'^2 -\lambda'\nu' + 2\frac{\nu' - \lambda'}{r}\right)\right],\label{fieq2233}\end{eqnarray}
where prime denotes derivative with respect to $r$, {$P$ is the isotropic pressure} and $\rho$ is the energy density.

At the outside of the fluid distribution, the spacetime is that of Schwarzschild, given by
\begin{eqnarray}
ds^2&=& \left(1-\frac{2M}{r}\right) dt^2 - \left(1-\frac{2M}{r}\right)^{-1}d{r}^2 \nonumber \\
&&\;\;\;\;\;\;\;\;\;\;\;\;\;\;\;\;\;\;\;\;\;\;
-{r^2} \left(d\theta^2 + \sin^2\theta d\phi^2 \right),\label{Schw_ext}
\end{eqnarray}
In order to match smoothly the two metrics above on the boundary surface $r=r_\Sigma$, we must require the continuity of the first and the second fundamental form across that surface (Darmois conditions). Then it follows 
\begin{equation}
e^{\nu_\Sigma}=1-\frac{2M}{r_\Sigma},\label{enusigma}
\end{equation}
\begin{equation}
e^{-\lambda_\Sigma}=1-\frac{2M}{r_\Sigma},\label{elambdasigma}
\end{equation}
\begin{equation}P_\Sigma=0,\label{PQ}
\end{equation}
where the subscript $\Sigma$ indicates that the quantity is evaluated at the boundary surface $\Sigma$.\noindent

Next, it will be useful to calculate the radial component of the conservation law\begin{equation}T^\mu_{\nu;\mu}=0, \label{dTmn}\end{equation}
where
\begin{equation}
T_{\mu\nu} = \left(\rho+P\right)u_\mu u_\nu - P g_{\mu\nu}, \label{T-}
\end{equation}
with
\begin{equation}
u^\mu=\left(e^{-\nu/2}, \, 0,\, 0, \, 0\right),\label{umu}
\end{equation}
where $u^\mu$ denotes the four velocity of the fluid.

After  simple calculations we get
\begin{equation}
P'=-\frac{\nu'}{2}\left(\rho+P\right),\label{Prpf}
\end{equation}
or using,
\begin{equation}
\nu' = \frac{2(m + 4 \pi P r^3)}{r \left(r - 2m\right)},
\label{nupri}
\end{equation}
we may also write
\begin{equation}
P'=-\frac{(m + 4 \pi P r^3)}{r \left(r - 2m\right)}\left(\rho+P\right),\label{Prpfii}
\end{equation}
which is the well known  Tolman--Oppenheimer--Volkoff equation, and 
where   the mass function $m(r)$, as usually is  defined by 
\begin{equation}
e^{-\lambda}=1-2m/r.\label{mass}
\end{equation}

\begin{figure}
\includegraphics[width=3.in,height=4.in,angle=0]{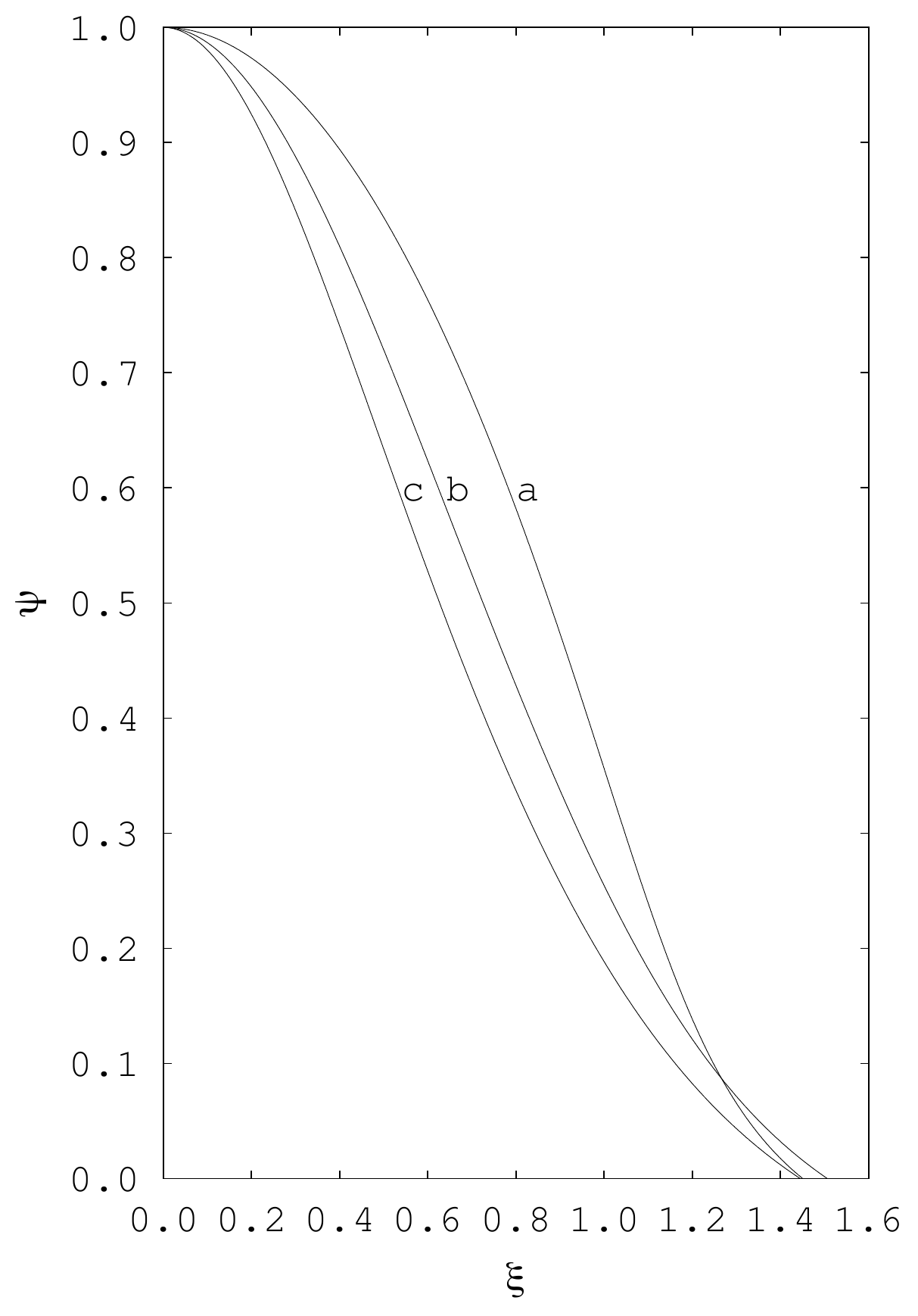}
\caption{$\psi$ as a function of $\xi$ for $\alpha=1$, $n=1$ and $h=0.5$ (curve a); $h=1.0$ (curve b); $h=1.5$ (curve c). Note the anomaly for $h=0.5$, that is, $\psi_0(\xi;h=1) > \psi_0(\xi;h=0.5)$ in some sector close to the surface $\xi_\Sigma$.}
\label{fig:II2}
\end{figure}
\begin{figure}
\includegraphics[width=3.in,height=4.in,angle=0]{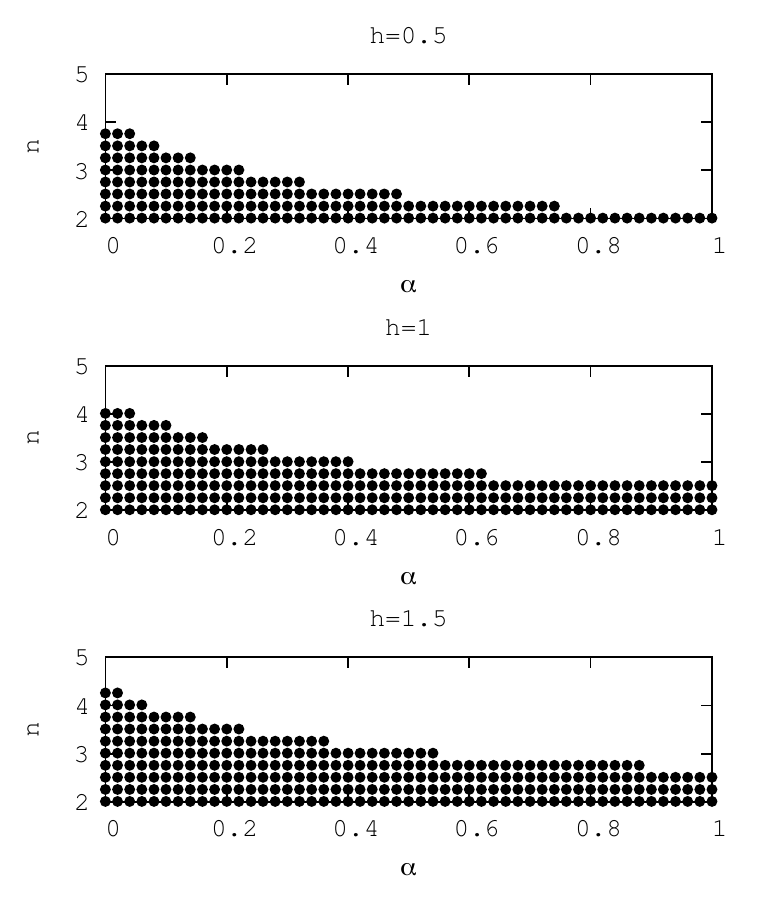}
\caption{Case II: Duplets $(\alpha,n)$ for bounded sources with $h$: 0.5$; 1.0$; $1.5$. Clearly $h>1$ favors the bounded configurations whereas  $h<1$ does not. Below $n=2$ (not shown) all the sources are bounded too.}
\label{fig:II8b}
\end{figure}
\begin{figure}
\includegraphics[width=3.in,height=4.in,angle=0]{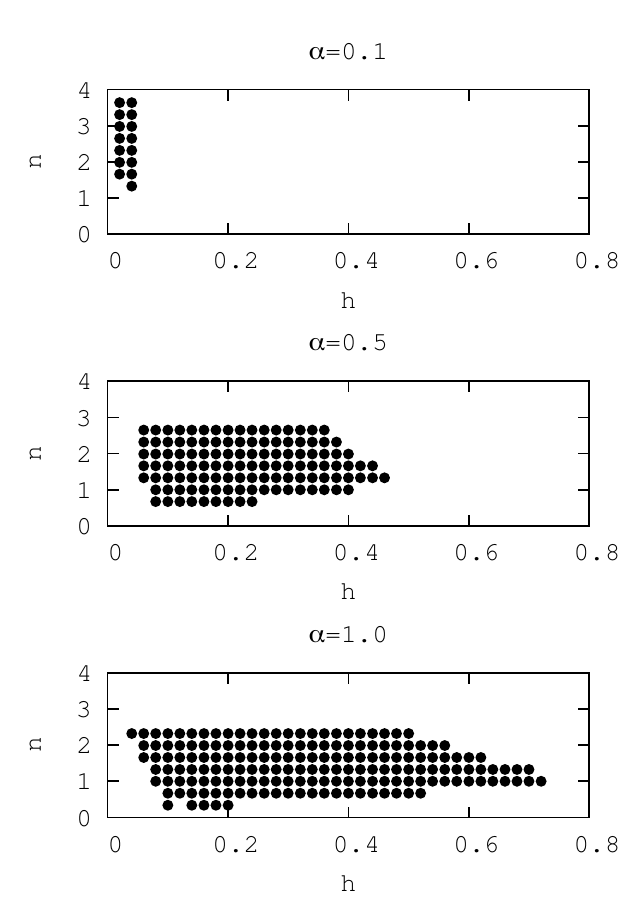}
\caption{Case II: Duplets $(h,n)$ for different $\alpha$ of bounded sources contained in figure \ref{fig:II8b}, which show anomalous behaviour like in figure \ref{fig:II2}.
To find this subset of anomalous models, we do a search of parameters for which the behaviour appears at least in some interior point of the distribution.} 
\label{fig:IIanoma}
\end{figure}

\subsection{The Relativistic Polytrope}
As it was mentioned in the Introduction the Newtonian polytrope is  characterized by (\ref{Pol}). However, when considering the polytropic equation of state within the context of general relativity, two different  possibilities arise, leading to the same equation (\ref{Pol}) in the Newtonian limit, namely:
\subsubsection{Case I}In this case the original polytropic equation of state, (\ref{Pol}),  is conserved,
see for example \cite{4c, 4b, 12}.

Now,  the first and the second law of thermodynamics may be written as
\begin{equation}
d(\frac{\rho + P}{\cal N})-\frac{dP}{\cal N}=Td(\frac{\sigma}{\cal N}) ,\label{1ley}\end{equation}
where $T$ denotes denote temperature, $\sigma$ is entropy per  unit of proper volume and $\cal N$ is the particle density, such that
\begin{equation}
\rho_0={\cal N} m_0.\label{n}\end{equation}

Then for an adiabatic process it follows
\begin{equation}
d(\frac{\rho}{\cal N})+Pd(\frac{1}{\cal N})=0 ,\label{1ley11}\end{equation}
which together with (\ref{Pol}) leads to
\begin{equation}
K\rho_{0}^{\gamma-2}=\frac{d(\rho/\rho_{0})}{d\rho_{0}} ,\label{Poladiabdif}\end{equation}
with

\begin{equation}\gamma=1+1/n .\label{gamma}\end{equation}
If $\gamma\neq 1$, (\ref{Poladiabdif}) may be easily integrated to give
\begin{equation}	\rho= C \rho_0 +  P/(\gamma-1) ,\label{densnueva}\end{equation}
where $C$ is a constant of integration.
In the non--relativistic limit we should have $\rho\rightarrow\rho_{0}$, and therefore $C=1$.

Thus, the polytropic equation of state amounts to
\begin{equation}	
\rho= \rho_0 +  P/(\gamma-1) ,\label{densnuevaI}\end{equation}
It is worth noticing that the familiar ``barotropic'' equation of state
\begin{equation}	
\rho=  P/(\gamma-1) ,\label{densnuevaII}\end{equation}
is a particular case of (\ref{densnueva}) with $C=0$. 

In the very special case $\gamma=1$, one obtains
\begin{equation}
K\rho_{0}^{-1}=\frac{d(\rho/\rho_{0})}{d\rho_{0}} ,\label{Poladiabdif3}\end{equation}
whose solution is
\begin{equation}	
\rho= P\log\rho_{0} +\rho_{0} C,\label{densgamma1}\end{equation}

or, if putting $C=1$ from the non--relativistic limit
\begin{equation}	\rho= P\log\rho_{0} +\rho_{0}.\label{densgamma11}\end{equation}
From now on we shall only consider the $\gamma\neq 1$ case.

Next, let us introduce the following variables
\begin{equation}
\alpha=P_c/\rho_{c},\quad r=\xi/A,  \quad A^2=4 \pi \rho_{c}/\alpha (n+1)\label{alfa},\end{equation}

\begin{equation}\psi_{0}^n=\rho_{0}/\rho_{0c},\quad v(\xi)=m(r) A^3/(4 \pi\rho_{c}),\label{psi}\end{equation}
where subscript $c$ indicates that the quantity is evaluated at the center. It is worth noticing that some of the variables defined above differ from those used in \cite{4b}.

Then the Tolman--Oppenheimer--Volkoff equation (\ref{Prpfii}) becomes
\begin{equation}
\xi^2 \frac{d\psi_{0}}{d\xi}\left[\frac{1-2(n+1)\alpha v/\xi}{(1-n\alpha)+(n+1)\alpha \psi_{0}}\right]+v+\alpha\xi^3 \psi_{0}^{n+1}=0,\label{TOV1}\end{equation}
and from the definition of mass function (\ref{mass}) and equation (\ref{fieq00}), we have
\begin{equation}
m'=4 \pi r^2 \rho\label{mprima}\end{equation}or
\begin{equation}
\frac{dv}{d\xi}=\xi^2 \psi_{0}^n (1-n\alpha+n\alpha\psi_{0}).\label{veprima}\end{equation}

Once again, recall that our definitions (\ref{alfa}) and (\ref{psi}) differ from Eqs. (15)--(18) in \cite{4b}, therefore (\ref{TOV1}) and (\ref{veprima}) also differ in form from Eqs. (19), (20) in \cite{4b}, although, of course they are absolutely equivalent.

The boundary of the surface of the sphere is defined  by $\xi=\xi_n$ such that $\psi_0(\xi_n)=0$ and the following boundary conditions apply:
\begin{equation}
\psi_{0}(\xi=0)=1,\,\,\, v(\xi=0)=0.
\label{boundary}
\end{equation}

Combining (\ref{TOV1}) and (\ref{veprima}) we obtain the generalized Lane--Emden equation for this case, which reads:
\begin{widetext}
\begin{equation}
a\frac{d^2\psi_{0}}{d\xi^2}+\frac{2}{\xi}\frac{d\psi_{0}}{d\xi}\left \{a-\alpha \xi(n+1)\left[\xi \psi^n_0(b-\alpha \psi_0)-\frac{v}{\xi^2}-\frac{a}{2b}\frac{d\psi_{0}}{d\xi}-\frac{b\xi\psi_0^n}{2}\right]\right \}+b \psi^{n}_0(3\alpha \psi_{0}+b-\alpha \psi_0)=0,\label{LEG1}
\end{equation}
\end{widetext}
\noindent where $$a\equiv 1-2(n+1)\alpha v/\xi$$ and $$b\equiv (1-n\alpha)+(n+1)\alpha \psi_{0}.$$
In the Newtonian limit ($\alpha\rightarrow0$), (\ref{LEG1}) becomes

\begin{equation}\frac{d^2\psi_{0}}{d\xi^2}+\frac{2}{\xi}\frac{d\psi_{0}}{d\xi}+\psi_{0}^n=0 \label{LNN},\end{equation}
which is the classical Lane--Emden equation.
\subsubsection{Case II}Another possibility consists in   assuming that the relativistic polytrope is defined by
\begin{equation}
P=K \rho^{1+1/n},\label{PolII}\end{equation}instead of (\ref{Pol}), see for example \cite{4a, 6, 13}.

In this case one obtains from (\ref{1ley11}) and (\ref{PolII})

{
\begin{equation}
\rho=\frac{\rho_0}{(1-K\rho_0^{1/n})^n}. \label{neqWB}
\end{equation}
Using (\ref{PolII}) and (\ref{alfa}) it can be easily demonstrated that if $\alpha$ is small enough, (\ref{neqWB}) becomes
\begin{equation}
\rho\approx \rho_0(1+nK\rho_0^{1/n}),\label{neqIIWB2}
\end{equation}
where only terms linear in $\alpha$ have been kept.
It should be observed that (\ref{neqIIWB2}) and (\ref{densnuevaI}) are identical, implying that both cases, I and II, differ in terms of order $\alpha^2$ and higher (obviously they coincide in the Newtonian limit).
}

Then introducing
\begin{equation}
\psi^n=\rho/\rho_{c},\label{psi2}\end{equation}
we obtain for  the  TOV equation
\begin{equation}\xi^2 \frac{d\psi}{d\xi}\left[\frac{1-2(n+1)\alpha v/\xi}{1+\alpha\psi}\right]+v+\alpha\xi^3 \psi^{n+1}=0, \label{TOV2}\end{equation}
and from (\ref{mprima})
\begin{equation}\frac{dv}{d\xi}=\xi^2 \psi^n.\label{veprima2}\end{equation}

From the two equations above, we can obtain the generalized Lane--Emden equation for this case, which reads
\begin{widetext}
\begin{equation}
a\frac{d^2\psi}{d\xi^2}+\frac{2}{\xi}\frac{d\psi}{d\xi}\left \{a-\alpha \xi^2(n+1) \psi^n+\frac{v(n+1)\alpha}{\xi}-\frac{d\psi}{d\xi}\frac{\xi a \alpha}{2c}+\frac{\xi^2 \alpha c(n+1)\psi^n}{2}\right \}+c \psi^{n}+3\alpha c\psi^{(n+1)}=0,\label{LEG11}\end{equation}
\end{widetext}
where $c\equiv1+\alpha \psi$.

Once again, in the Newtonian limit   ($\alpha\rightarrow0$), the Lane--Emden equation (\ref{LNN}) is recovered in this case too, as it should be. Obviously  equations of state in both cases differ from each other, especially in the highly relativistic regime, and therefore resulting models will also differ. 

We shall next generalize the scheme above for the case when the pressure is no longer isotropic.

\section{The polytrope for anisotropic fluids}
If we allow the principal stresses to be unequal, then the energy--momentum  tensor in the canonical form reads
\begin{equation}
T^{\mu}_{\nu}=\rho u^{\mu}u_{\nu}-  P
h^{\mu}_{\nu}+\Pi ^{\mu}_{\nu},
\label{24'}
\end{equation}
where  $P$ is  the isotropic pressure, and $\Pi_{\mu \nu}$ the anisotropic pressure tensor,
with 
\begin{equation}
h^{\mu}_{\nu}=\delta^{\mu}_{\nu}-u^{\mu}u_{\nu},\quad \Pi
^{\mu}_{\nu}=\Pi(s^{\mu}s_{\nu}+\frac{1}{3}h^{\mu}_{\nu}),
\label{nt}
\end{equation}
and  $s^\mu$ is defined as
\begin{equation}
s^{\mu}=(0, e^{-\frac{{\lambda}}{2}},0,0),
\end{equation}
with the  properties
$s^{\mu}u_{\mu}=0$,
$s^{\mu}s_{\mu}=-1$.

It is immediate to see that:

\begin{equation} \label{jc10} \rho = T_{\alpha \beta} u^\alpha u^\beta, \;  \end{equation} 
\begin{equation}\label{jc11} P = -\frac13 h^{\alpha \beta} T_{\alpha \beta}, \;\; \Pi_{\alpha \beta} = h_\alpha^\mu h_\beta^\nu \left(T_{\mu\nu} + P h_{\mu\nu}\right). \end{equation}

For our purposes in this work, it would be more convenient to introduce the following  two auxiliary variables ($P_r$ and $P_\perp$) by:  

\begin{equation}
P_r=s^\alpha s^\beta T_{\alpha \beta},\qquad P_\perp=K^\alpha K^\beta T_{\alpha \beta},
\end{equation}
where $K^\alpha$ is a  unit spacelike vector (orthogonal to $u^\alpha$ and $s^\alpha$).

In terms of the above variables, we have
\begin{equation}
\Pi=P_r-P_\perp ; \qquad P=\frac{P_r+2 P_\perp}{3},
\label{ns}
\end{equation}
from where the physical meaning of $P_r$ and $P_\perp$ becomes evident, and the energy momentum can be written under the form

\begin{equation}
T_{\mu\nu} = \left(\rho+P_\bot\right)u_\mu u_\nu - P_\bot g_{\mu\nu} + \left(P_r-P_\bot\right)s_\mu s_\nu. \end{equation}

From these last  expressions it a simple matter to prove that the hydrostatic equilibrium  equation now reads
\begin{equation}
P'_r=-\frac{\nu'}{2}\left(\rho+P_r\right)+\frac{2\left(P_\bot-P_r\right)}{r},\label{Prp}
\end{equation}
where $P_r$ and $P_{\bot}$ will be hereafter  called the radial and tangential pressures, respectively. This is the  generalized Tolman-Opphenheimer-Volkoff equation for anisotropic matter. 
Alternatively, using 
\begin{equation}
\nu' = 2 \frac{m + 4 \pi P_r r^3}{r \left(r - 2m\right)},
\label{nuprii}
\end{equation}
we may write
\begin{equation}
P'_r=-\frac{(m + 4 \pi P_r r^3)}{r \left(r - 2m\right)}\left(\rho+P_r\right)+\frac{2\left(P_\bot-P_r\right)}{r}.\label{ntov}
\end{equation}
Based on the considerations of the previous section, we shall consider the following two  cases  to extend the polytropic equation of state to anisotropic matter.
\begin{enumerate}
\item \begin{equation}P_r=K\rho_0^{\gamma}=K\rho_0^{1+1/n} \label{I}.\end{equation}
\item \begin{equation}P_r=K\rho^{\gamma}=K\rho^{1+1/n} \label{II}.\end{equation}
\end{enumerate}

We shall next proceed to analyze each case in detail.
\subsection{Case I}
Assuming (\ref{I}), then with the help of (\ref{1ley}) we obtain 
\begin{equation}	
\rho= \rho_0 +  P_r/(\gamma-1) .\label{densnuevaIanis}\end{equation}
Then repeating the same procedure as in the isotropic case  we get
\begin{widetext}
\begin{eqnarray}
\xi^2 \frac{d\psi_{0}}{d\xi}\left[\frac{1-2(n+1)\alpha v/\xi}{(1-n\alpha)+(n+1)\alpha \psi_{0}}\right]+v+\alpha\xi^3 \psi_{0}^{n+1}+\frac{2\Delta \psi_0^{-n}\xi}{P_{rc}(n+1)} \left[\frac{1-2\alpha (n+1)v/\xi}{(1-n\alpha)+(n+1)\alpha \psi_0}\right]=0,\label{TOV1anis_WB}
\end{eqnarray}
\end{widetext}
where now $\alpha=P_{cr}/\rho_c$ and $\Delta=-\Pi=P_\bot-P_r$.

On the other hand we obtain from the mass function definition (\ref{mass}) the same equation (\ref{veprima}) as in the isotropic case.

Combining (\ref{TOV1anis_WB}) and (\ref{veprima}) we are led to the generalized Lane-Emden equation for this case
\begin{widetext}
{
\begin{eqnarray}
&&a\frac{d^2\psi_{0}}{d\xi^2}+\frac{2}{\xi}\frac{d\psi_{0}}{d\xi}\left \{a-\alpha \xi(n+1)\left[\xi \psi^n_0(b-\alpha \psi_0)-\frac{v}{\xi^2}-\frac{a}{2b}\frac{d\psi_{0}}{d\xi}-\frac{b\xi\psi_0^n}{2}\right] -\frac{\xi^2}{(n+1)P_{rc}}\left[\psi_0^{-n}\Delta\frac{a}{b}\right]\left[\frac{(n+1)\alpha}{b}+\frac{n}{\psi_0}\right]\right\}\nonumber\\
&+&b \psi^{n}_0(3\alpha \psi_{0}+b-\alpha \psi_0)
+\frac{2}{(n+1)P_{rc}}\left[\psi_0^{-n}\Delta\xi\frac{a}{b}\right]\left\{\frac{1}{\Delta}\frac{d\Delta}{d\xi}+ \frac{1}{\xi} + \frac{2(n+1)\alpha}{a}\left[-\xi\psi_0^n[1+n\alpha(\psi_0-1)]+\frac{v}{\xi^2}\right]\right\}=0,\label{LEG1_anis}
\end{eqnarray}}
\end{widetext}
It is a simple matter to check that the equation above reduces to Eq. (20) in \cite{HB} in the Newtonian limit (taking care of the changes in notation).

\subsection{Case II}
In this case the assumed equation of state is (\ref{II}), 
then the TOV equation becomes
\begin{widetext}

\begin{eqnarray}
\xi^2 \frac{d\psi}{d\xi}\left[\frac{1-2(n+1)\alpha v/\xi}{1+\alpha \psi}\right]+v+\alpha\xi^3 \psi^{n+1}+\frac{2\Delta \psi^{-n}\xi}{P_{rc}(n+1)} \left[\frac{1-2\alpha(n+1)v/\xi}{1+\alpha \psi}\right]=0,\label{TOV2anis_WB}
\end{eqnarray}
\end{widetext}
and from the definition of the mass function (\ref{mass}), we obtain (\ref{veprima2}).

Once again, the combination of (\ref{TOV2anis_WB}) and (\ref{veprima2}) leads to the generalized Lane-Emden equation for this case which reads:
\begin{widetext}

\begin{eqnarray}
&&a\frac{d^2\psi}{d\xi^2}+\frac{2}{\xi}\frac{d\psi}{d\xi}\left \{a-\alpha \xi^2(n+1) \psi^n+\frac{v(n+1)\alpha}{\xi}-\frac{d\psi}{d\xi}\frac{\xi a \alpha}{2c}+\frac{\xi^2 \alpha c(n+1)\psi^n}{2}-\frac{\xi^2}{(1+n)P_{rc}}\left[\Psi^{-n}\Delta\frac{a}{c}\right]\left[\frac{n}{\psi}+\frac{\alpha}{c}\right]\right\}\nonumber\\
&+&c \psi^{n}+3\alpha c\psi^{(n+1)}
+ \frac{2}{(1+n)P_{rc}}\left[\psi^{-n}\Delta \xi \frac{a}{c}\right]\left\{\frac{1}{\Delta}\frac{d\Delta}{d\xi}+\frac{1}{\xi}+\frac{2(n+1)\alpha}{a}\left[-\xi\psi^n+\frac{v}{\xi^2}\right]\right\}=0,\label{LEG11}
\end{eqnarray}
\end{widetext}
In the Newtonian limit we recover the Eq. (20) in \cite{HB}.

{As mentioned before, it is obvious that in order to proceed further with the modeling of a compact object, we need to prescribe the specific anisotropy of the problem ($\Delta$). Such information, of course, depends on the specific physical problem under consideration. Here we shall not follow that direction; instead, we shall assume an ansatz already used in the modeling of relativistic anisotropic  stars \cite{chew81,chew82}, whose main virtue (besides its simplicity) is the fact that the obtained models are continuously connected with the isotropic case, thus allowing to compare both cases, and thereby to illustrate the influence of the anisotropy on the structure of the object.}
\begin{figure}
\includegraphics[width=3.in,height=4.in,angle=0]{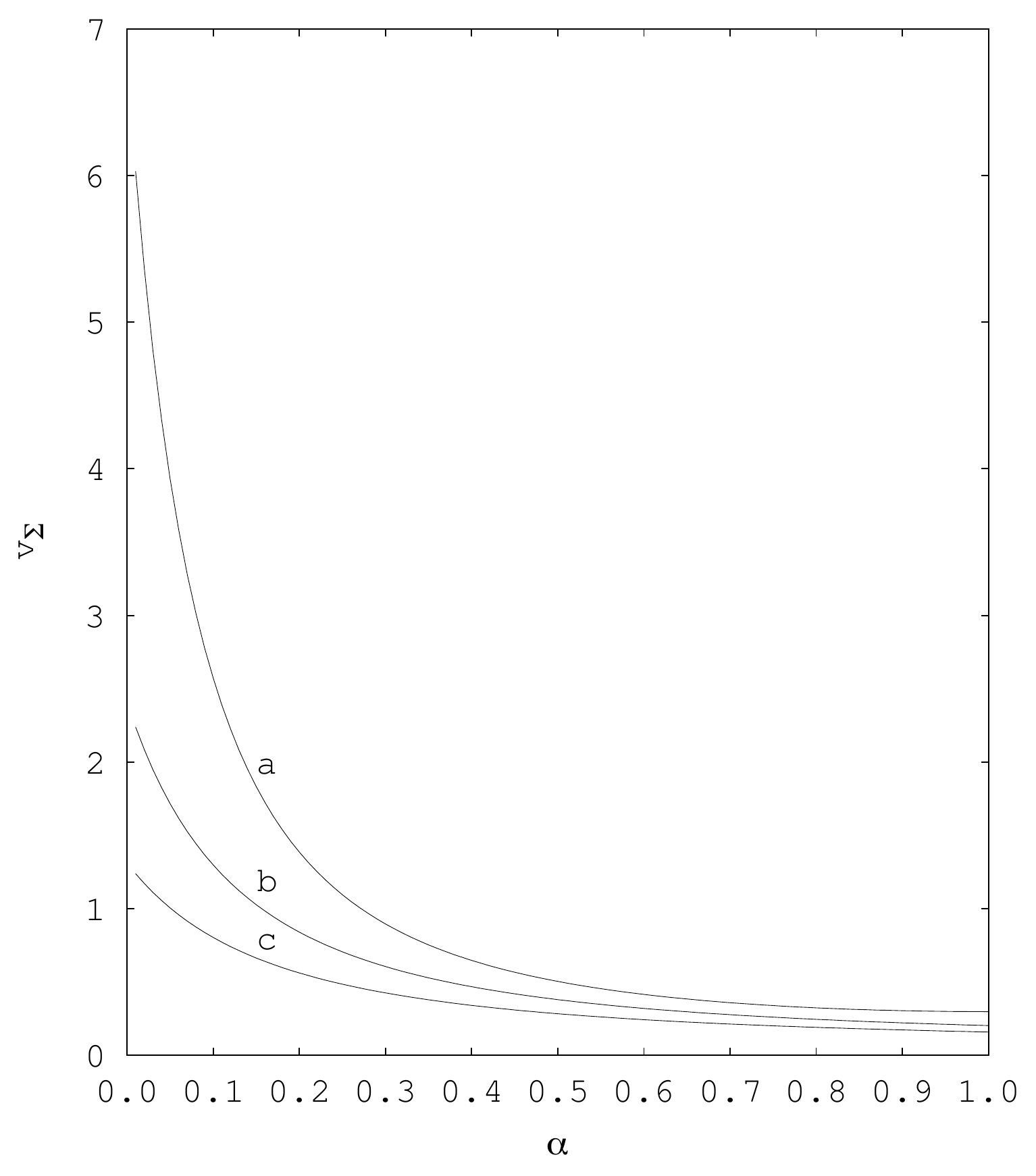}
\caption{Case II: $v_\Sigma$ as a function of $\alpha$ for $n=2$ and $h=0.5$ (curve a); $h=1.0$ (curve b); $h=1.5$ (curve c). For a wide range of $(n,h)$ this figure is qualitatively representative.}
\label{fig:II4}
\end{figure}
\begin{figure}
\includegraphics[width=4.in,height=6.in,angle=0]{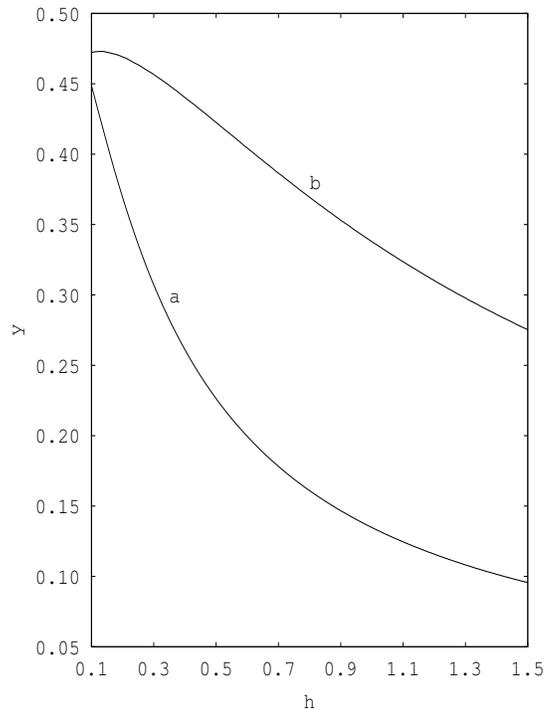}
\caption{Case I: Potential at the surface $y$ as a function of the anisotropy parameter, 
$h$ for pairs $(n,\alpha)$: $(1.0,0.1)$ (curve a); $(0.5,1)$ (curve b).}
\label{fig:surpotI}
\end{figure}
\begin{figure}
\includegraphics[width=4.in,height=6.in,angle=0]{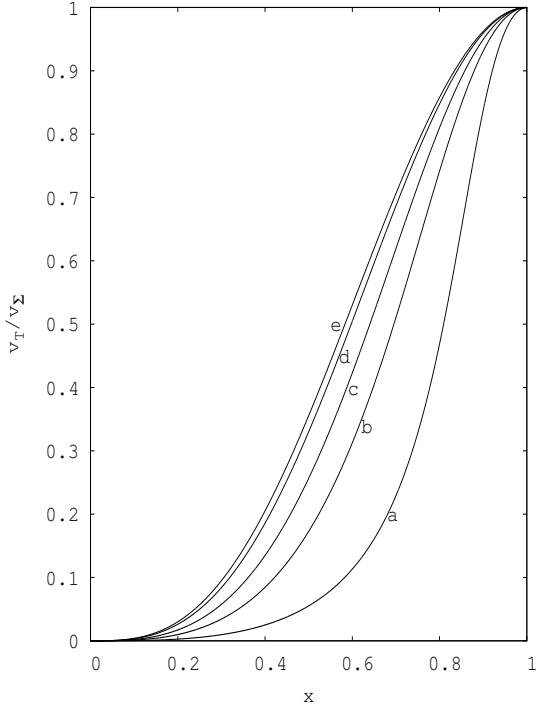}
\caption{Case I: $v_T/v_\Sigma$ as a function of $x$ for $n=1$, $\alpha=0.1$, and different values of $h \,(y)$: $0.15 \,(\approx 0.41)$ (curve a); $0.3 \,(\approx 0.31)$ (curve b); $0.5 \,(\approx 0.23)$ (curve c), $1.0 \,(\approx 0.13)$ (curve d), $1.5 \,(\approx 0.1)$ (curve e). Values of $y$ are read off figure \ref{fig:surpotI}. The value of $h=0.1$ corresponds to $y = 0.448 \,(> 4/9)$.}
\label{fig:I15}
\end{figure}
\begin{figure}
\includegraphics[width=4.in,height=6.in,angle=0]{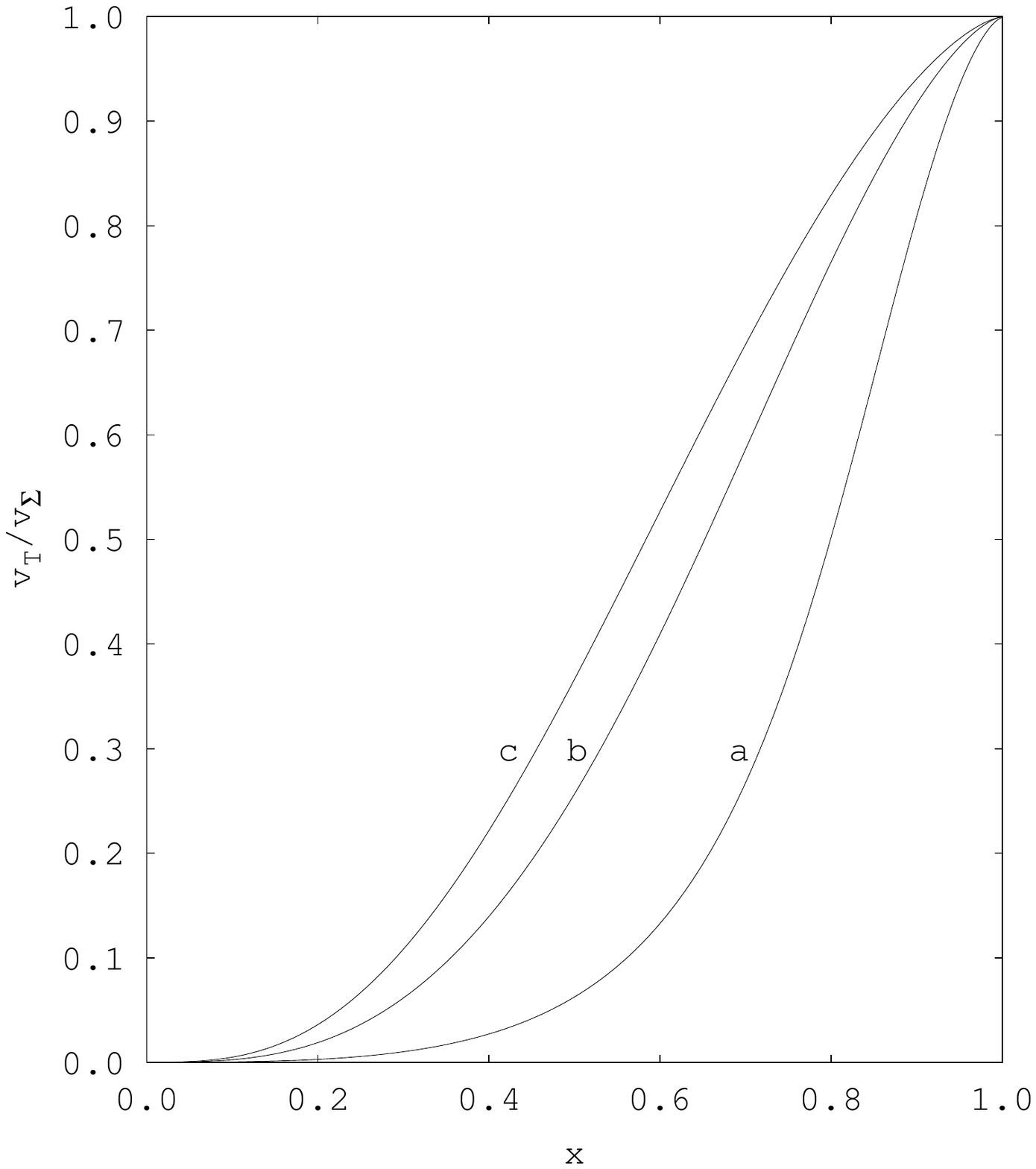}
\caption{Case I: $v_T/v_\Sigma$ as a function of $x$ for $n=0.5$, $\alpha=1.0$, and different values of $h \,(y)$: $0.5 \,(\approx 0.42)$ (curve a), $1.0 \,(\approx 0.34)$ (curve b), $1.5 \,(\approx 0.28)$ (curve c). Values of $y$ are read off figure \ref{fig:surpotI}.}
\label{fig:I17}
\end{figure}
\begin{figure}
\includegraphics[width=4.in,height=6.in,angle=0]{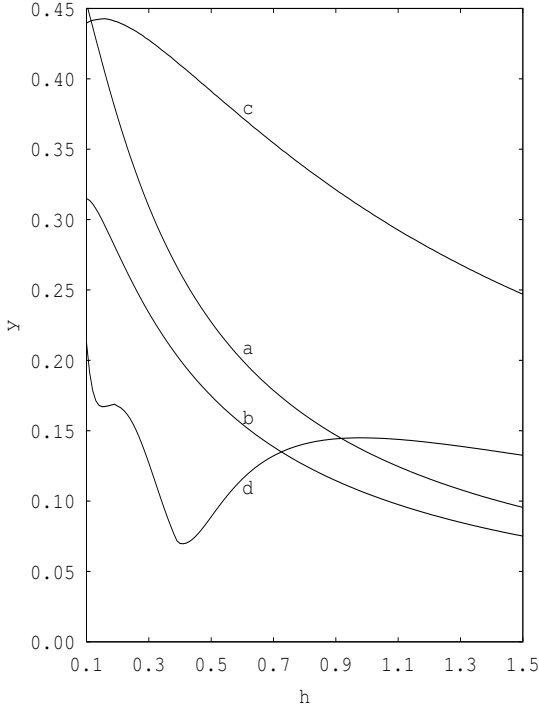}
\caption{Case II: Potential at the surface $y$ as a function of the anisotropy parameter, 
$h$ for pairs $(n,\alpha)$: $(1.0,0.1)$ (curve a); $(2.0,0.1)$ (curve b), $(1.0,1.0)$ (curve c), $(2.0,1.0)$ (curve d)}
\label{fig:surpotII}
\end{figure}
\begin{figure}
\includegraphics[width=4.in,height=6.in,angle=0]{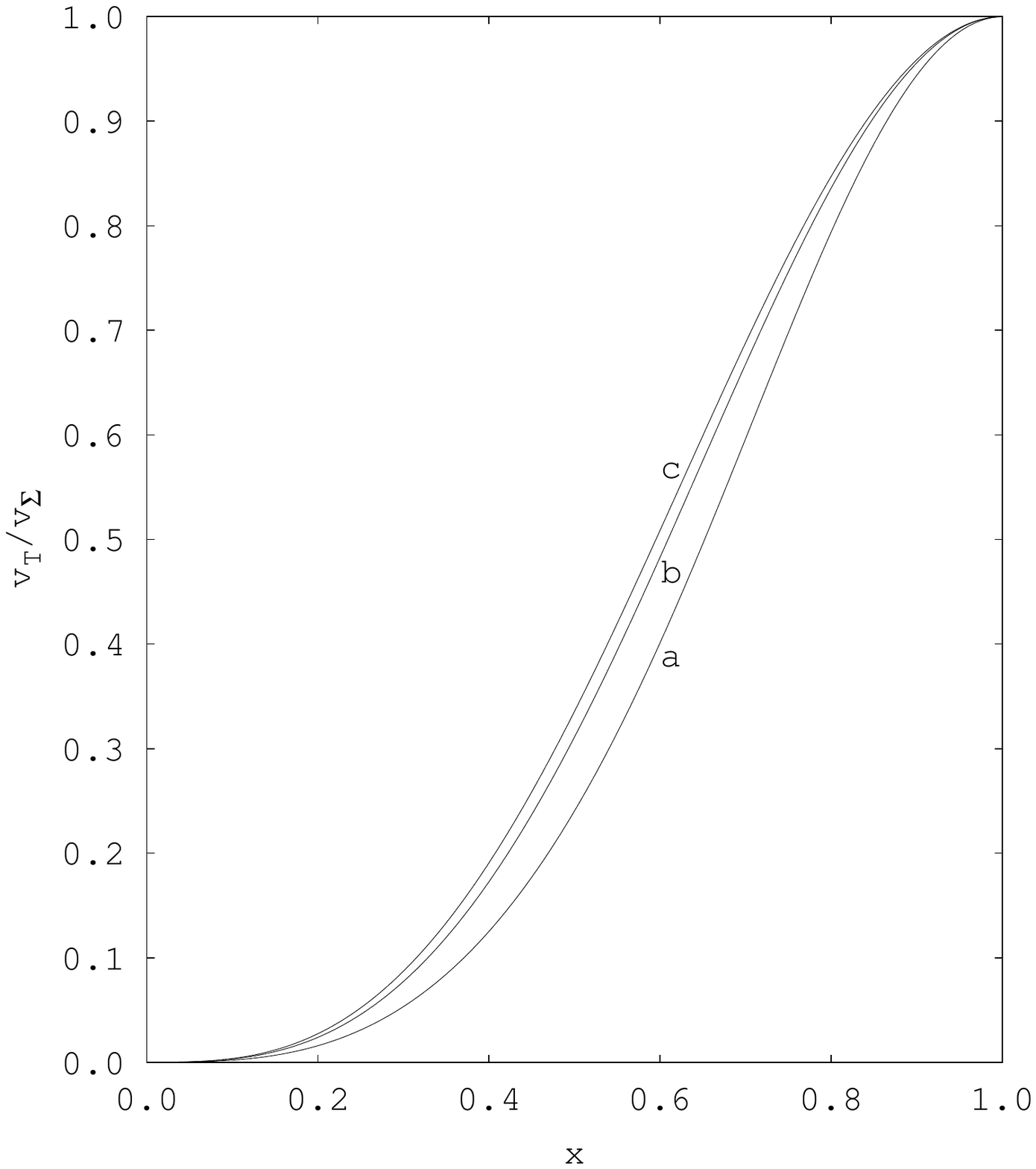}
\caption{Case II: $v_T/v_\Sigma$ as a function of $x$ for $n=1$, $\alpha=0.1$, and different values of $h \,(y)$: $0.5 \,(\approx 0.23)$ (curve a), $1.0 \,(\approx 0.13)$ (curve b), $1.5 \,(\approx 0.1)$ (curve c). Values of $y$ are read off figure \ref{fig:surpotII}.}
\label{fig:II15}
\end{figure}
\section{Modeling anisotropic polytropes}
In order to obtain specific models to illustrate our method, we shall  here adopt the   heuristic  procedure used in Ref. \cite{chew81}, which allows one to obtain solutions for anisotropic matter from known solutions for isotropic matter.
The procedure mentioned above may be summarized as follows (see \cite{chew81}) for details:
\begin{itemize}
\item Let 
\begin{equation}
\Delta=Cf(P_r,r)(\rho+P_r)r^N,
\label{pr1}
\end{equation}
where $C$ is a parameter which measures the anisotropy, and  the function $f$ and
the number $N$ are to be specific for each model. 
\item Assume 
\begin{equation}
f(P_r,r)r^{N-1}=\frac{\nu^{\prime}}{2}.
\label{pr2}
\end{equation}
\item With the assumption above,  (\ref{Prp}) becomes
\begin{equation}
\frac{dP_r}{dr}=-h(\rho +P_r)\frac{\nu^{\prime}}{2},
\label{pr2}
\end{equation}
where $h=1-2C$.  For simplicity, we assume $h$ to be constant throughout the sphere, which of course does not imply the constancy of either pressure. 
\end{itemize}
Then we obtain for the case I the following two equations
\begin{equation}
\xi^2 \frac{d\psi_{0}}{d\xi}\left[\frac{1-2(n+1)\alpha v/\xi}{(1-n\alpha)+(n+1)\alpha \psi_{0}}\right]+h(v+\alpha\xi^3 \psi_{0}^{n+1})=0,\label{TOV1h}\end{equation}
and 
\begin{equation}
\frac{dv}{d\xi}=\xi^2 \psi_{0}^n (1-n\alpha+n\alpha\psi_{0}),\label{veprimah}\end{equation}
whereas for the case II, the corresponding equations read
\begin{equation}\xi^2 \frac{d\psi}{d\xi}\left[\frac{1-2(n+1)\alpha v/\xi}{1+\alpha\psi}\right]+h(v+\alpha\xi^3 \psi^{n+1})=0, \label{TOV2h}\end{equation}
and 
\begin{equation}
\frac{dv}{d\xi}=\xi^2 \psi^n.\label{veprima2h}
\end{equation}

The above equations shall be integrated numerically, with the corresponding boundary conditions.

It will  be useful to  calculate the Tolman mass, which is   a measure of the active gravitational mass (see  \cite{hbds02} and references therein), defined by
\begin{equation}
m_T=\frac{1}{2} r^2e^{(\nu-\lambda)/2} \nu'.
\end{equation}

Alternatively we can calculate the Tolman mass from the expression \cite{14}
\begin{equation}
m_T=e^{(\nu+\lambda)/2} (m+4\pi P_r r^3).
\label{mt1}
\end{equation}
The functions  $\lambda$, $m$ and $P_r$ in the above expression, are obtained directly by integration of (\ref{TOV1h}) and (\ref{veprimah}) for the case I, and  (\ref{TOV2h}) and (\ref{veprima2h})  for the  case II. Thus we only need an expression for $\nu$, which can be obtained as follows, for either case.
Let us first consider the case I. Then from (\ref{I}), (\ref{psi}) and (\ref{densnuevaIanis}), we may write
\begin{equation}
P_r=K\rho_{0c}^{1+1/n}\psi_0^{(n+1)}
\label{mt2}
\end{equation}
and
\begin{equation}
\rho=\rho_{0c}\psi_0^n(1+nk\rho_{0c}^{1/n}\psi_0),
\label{mt3}
\end{equation}
with the help of these two expressions we may write Eq. (\ref{pr2}) in the form
\begin{equation}
2(n+1)\beta d\psi_0+hd\nu \left[1+\beta\psi_0 (n+1)\right]=0,
\label{mt4}
\end{equation}
where the parameter $\beta \equiv K\rho_{0c}^{1/n}$ is related to $\alpha$ by 
\begin{equation}
\beta=\frac{\alpha}{1-n\alpha}.
\label{mt5}
\end{equation}
Equation (\ref{mt4}) can be easily integrated to obtain
\begin{equation}
e^\nu=\frac{C}{\left[1+(n+1)\beta \psi_0\right]^{2/h}},
\label{mt6}
\end{equation}
where $C$ is a constant of integration.
Next, using the fact that at the center ($r=0$) $\psi_0=1$, we may write (\ref{mt6}) as
\begin{equation}
e^\nu=e^{\nu_c}\frac{\left[1+(n+1)\beta\right]^{2/h}}{\left[1+(n+1)\beta \psi_0\right]^{2/h}}.
\label{mt7}
\end{equation}
Finally, using (\ref{enusigma}) and the fact that the radial pressure vanishes at the boundary surface  we obtain
\begin{equation}
e^\nu=\left(1-\frac{2M}{r_\Sigma}\right)\left[1+(n+1)\beta \psi_0\right]^{-2/h}.
\label{mt8}
\end{equation}

Thus, the (dimensionless) Tolman mass can be written as
\begin{widetext}
\begin{equation}
v_T=(v+\alpha\xi^3\psi_0^{1+n})(a_\Sigma/a)^{1/2}\left[\frac{1-n\alpha}{1-n\alpha+\alpha(1+n)\psi_0}\right]^{1/h},
\end{equation}
\end{widetext}
where
$$v_T=\frac{m_T A^3}{4\pi\rho_c}.$$

{For the case II, we follow a similar procedure. 
Thus, using (\ref{II}) and (\ref{psi2}), we transform (\ref{pr2}) into 
\begin{equation}
2(n+1)\alpha d\psi+h(1+\alpha \psi)d\nu=0 ,
\label{mt9}
\end{equation}
which after integration  produces
\begin{equation}
e^\nu=\left(1-\frac{2M}{r_\Sigma}\right)\left(1+\alpha \psi\right)^{-2(n+1)/h},
\label{mt10}
\end{equation}
where boundary conditions have been used}. {In this case the Tolman mass reduces to
\begin{equation}
v_T=(v+\alpha\xi^3\psi^{1+n})(a_\Sigma/a)^{1/2}(1+\alpha\psi)^{-(1+n)/h}.
\end{equation}

In order to see how the Tolman mass distributes through the sphere in the process of  contraction (slow and adiabatic), it would be convenient to introduce the following dimensionless variables:
\begin{equation}
x=\frac{r}{r_\Sigma}=\frac{\xi}{\tilde A},\qquad y=\frac{M}{r_\Sigma},\qquad \tilde m=\frac{m}{M},\qquad \tilde A=r_\Sigma A.
\label{nm1}
\end{equation}
In terms of the above variables the Tolman  mass for the two cases I and II reads
\begin{widetext}
\begin{equation}
v_T=(v+\alpha x^3 \tilde A^3\psi_0^{1+n})\left[\frac{x(1-2y)}{x-2\alpha(n+1)v/\tilde A}\right]^{1/2}\left[\frac{1-n\alpha}{1-n\alpha+\alpha(1+n)\psi_0}\right]^{1/h}
\label{nmII}
\end{equation}
\end{widetext}
and 
\begin{widetext}
\begin{equation}
v_T=(v+\alpha x^3 \tilde A^3\psi^{1+n})\left[\frac{x(1-2y)}{x-2\alpha(n+1)v/\tilde A}\right]^{1/2}\left[1+\alpha \psi\right]^{-(1+n)/h},
\label{nmIII}
\end{equation}
\end{widetext}
respectively.
{Observe that $\tilde A=\xi_\Sigma$, therefore
\begin{equation}
y=\alpha(n+1) \frac{v_\Sigma}{\xi_\Sigma}.
\label{nmIV}
\end{equation}
\begin{figure}
\includegraphics[width=4.in,height=6.in,angle=0]{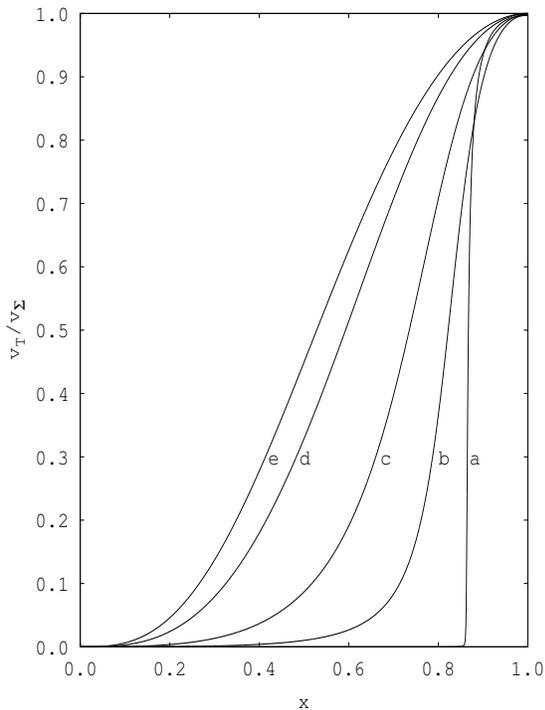}
\caption{Case II: $v_T/v_\Sigma$ as a function of $x$ for $n=1$, $\alpha=1.0$, and different values of $h\,(y)$: $0.1 \,(\approx 0.44)$ (curve a), $0.3 \,(\approx 0.43)$ (curve b),$0.5 \,(\approx 0.39)$ (curve c), $1.0 \,(\approx 0.31)$ (curve d), $1.5 \,(\approx 0.25)$ (curve e). Values of $y$ are read off figure \ref{fig:surpotII}.}
\label{fig:II17}
\end{figure}
\begin{figure}
\includegraphics[width=4.in,height=6.in,angle=0]{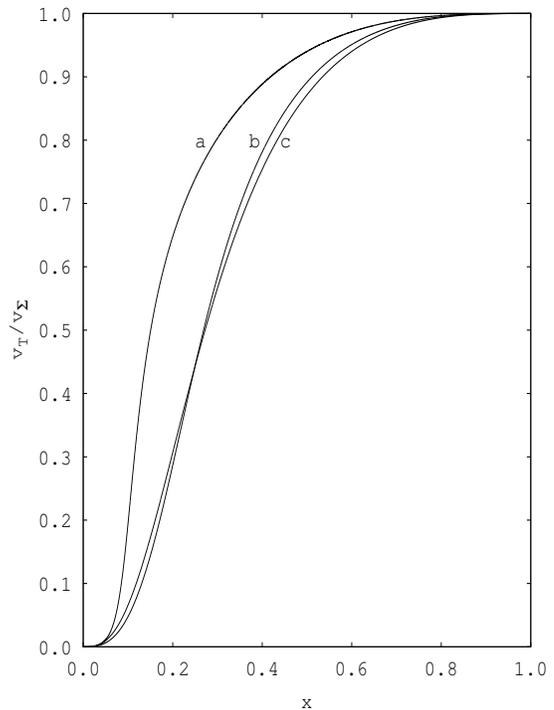}
\caption{Case II: $v_T/v_\Sigma$ as a function of $x$ for $n=2$, $\alpha=1.0$, and different values of $h(y)$: $0.5\,(\approx 0.09)$ (curve a), $1.0 \,(\approx 0.14)$ (curve b), $1.5 \,(\approx 0.13)$ (curve c). Values of $y$ are read off figure \ref{fig:surpotII}.}
\label{fig:II24}
\end{figure}

This means that $y$ is not independent of the anisotropy parameter $h$. In other words, $y$ is constant for a given pair $(n,\alpha)$ if $v_\Sigma/\xi_\Sigma=$ constant, and this is possible only for each $h$. One potential at the surface $y$ corresponds univocally to each anisotropic and relativistic polytrope.}

The full set of equations deployed above  has been integrated for both cases (I and II), and a wide range of values of different parameters. In what follows we analyze the most relevant results obtained from a selection of  the whole set of obtained models.

In the case I, Figure 1 shows  the integration of (\ref{TOV1h}) and (\ref{veprimah})  for the values of the parameters indicated in the figure legend. However as mentioned there, the ensuing  qualitative behaviour, namely, larger values of $\psi_0$ for smaller values of $h$ (everywhere throughout the sphere), is maintained for a wide range of values of  the triplet $(\alpha, n, h)$. There exists though, some range of values of the parameters, for which such a behaviour is disrupted, as indicated in Fig. 2, where $\psi_0$ close to the boundary surface is larger for $h=1$ than for $h=0.5$. We shall comment further on this ``anomaly'' later on.

In the Newtonian polytropes  there exist bounded configurations for a specific range of the parameter $n$, whereas  for  the general relativistic polytropes of an isotropic fluid, bounded configurations  are also restricted by the value of the parameter $\alpha$. In our case, as it should be expected, the existence of bounded configurations is restricted by the triplet  $(\alpha, n, h)$, as indicated in Figs. 3 and 4.

Finally, always for  the case I, in Fig. 5 we plot the total mass function ($v_\Sigma$) as function of the relativistic parameter $\alpha$ for  different values of $h$, and $n=2$. The exhibited behaviour, larger values of $v_\Sigma$ for smaller values of $h$, is representative for a wide range of  $(n, h)$. 

The case II is  described by the integration of (\ref{TOV2h}) and (\ref{veprima2h}). Results are exhibited in figures 6--10, and besides quantitative differences, the analysis of these figures is essentially the same as in the case I.

Next,  we have plotted the ``surface potential'' ($y$) for both cases, in figures 11 and 14. This is  a relevant variable since it measures the compactness of the configuration. For the case I (Fig.  11) we observe that the degree of compactness monotonically decreases with $h$. For the case II (Fig. 14) the situation is essentially the same (decreasing of $y$ with increasing $h$), except for some range of values of $h$ in the ultrarelativistic case and $n=2$ (curve d).  At any rate, maximal values of $y$ correspond to minimal values of $h$ in the four curves. 

It should be emphasized that the relationship between the maximal values of $y$  (maximal surface redshift) and the local anisotropy of pressure, has been discussed in great detail in the past (see \cite{mx1, mx2, mx3, mx5, mx6, mx4, mx7} and references therein). The explanation for such interest  is easily understood, if we recall that the surface redshift is an observable variable, which thereby might provide information about the structure of the source, in particular about its degree of anisotropy. As mentioned before each anisotropic polytropic model is characterized by a unique $y$ (see (\ref{nmIV})). The polytropic models favoring higher redshifts are clearly exhibited in the figures 11 and 14.

The correspondence mentioned above, between $y$ and $h$, suggests an increasing of the stability of the models with a decreasing $h$, except for the cases where such a correspondence is broken. In order to delve deeper into this interesting feature of the models, we have  investigated the behaviour of the Tolman mass within the sphere.

Figure 12 displays the Tolman mass (normalized by the total mass), for the case I, as function of $x$ for the selection of values of the parameters indicated in the legend. As we move from the less compact configuration (curve e) to the more compact one (curve a), the Tolman mass tends to concentrate on the outer regions of the sphere. This behaviour persists for a wide range of values of the parameters (see Fig. 13). Now in order to better understand this effect, we recall that we can regard the slow, adiabatic evolution of a sphere, as a sequence of  spheres in equilibrium. Thus in the process of contraction we move from $e$ to $a$, i.e, the system shifts to more compact configurations with smaller values of $h$ and with the Tolman mass decreasing in the inner regions. In other words as the sphere gets more compact, equilibrium configurations correspond to smaller values of $h$, which in turn produce a sharper effect of migration of Tolman mass towards the boundary surface.  In terms of stability, and keeping in mind the physical meaning of the Tolman mass, it may be said that more stable configurations correspond to  smaller values of $h$, since these are associated to a sharper reduction of the  active gravitational mass in the inner regions, thereby providing a clear physical picture  of the described scenario. It is worth mentioning that this effect was observed  some  years ago \cite{HS} in the case of the Florides solution \cite{F}.

For the case II the situation is essentially the same, as shown in Fig. 15, although with an important  difference. Indeed, for the values of the parameters considered in Fig. 15 and for a wide range of values, the behaviour of the Tolman mass is basically the same as in the case I. However  for the values of Fig. 16 we observe a deviation from that behaviour, in a small region close to the boundary surface, where the Tolman mass is larger for smaller values of $h$ (curve $a$), than for larger ones (curves $b$ and $c$). Again, speaking in terms of stability, it appears that for the case of figure 16, the stability is enhanced by decreasing of $h$, in the inner regions of the sphere, while  the opposite seems to happen in the outer layers. 
An extreme example of the effect mentioned above is shown in Fig. 17. As we can see, the curve $a$ corresponding to the minimal value of $h$, describes a configuration with the maximal value of the Tolman mass inside the sphere. As the adiabatic collapse proceeds, we move into the  curve $c$ corresponding to the maximal value of $h$, and from thereof, to the curve $b$ corresponding to the isotropic case. Thus the less stable configuration in this example is that corresponding to $a$. On the other hand the case $b$ is more stable than $c$ in some inner region, while the opposite happens in the outer layers.

So, it appears that the  stability  of the ``core'' and the ``envelope''  may respond differently to different degrees of anisotropy (different $h$). This fact was already pointed out, and discussed in some detail for the Bowers--Liang solution \cite{BL}, in \cite{HRW}. Here we have arrived at the same result  by analyzing the behaviour of the Tolman mass, whereas in \cite{HRW} such a peculiar effect,  was  obtained  by studying the behaviour of the adiabatic index.

{Finally, the following two comments are in order:
\begin{itemize}
 \item For both cases I and II we have found that for some selection of values of the parameters, the behaviour of functions $\psi_0$ and $\psi$, close to the boundary surface, deviates from the ``standard'' behaviour depicted in Figs. 1 and 6. This ``anomalous'' behaviour   is  shown in Figs. 2 and 7. 
On the other hand, another ``anomalous'' behaviour in the stability (Tolman mass) was found for the case II, close to the boundary (in some range of values of the parameters). Both behaviours occur for  similar values of the parameters. However, while the former appears in both cases,  the latter was only found  for the case II. Therefore the possible link between both kinds of anomalous behaviours, if any,  has not been established.
\item The fact that some features of our models (e.g. the different stability response to different degrees of anisotropy) are shared by a completely different  kind of anisotropic models, suggests that such features might be common to all, or at least to  a wide family of, anisotropic compact stars.
\end{itemize}
}
\section{conclusions}
We have presented the general framework for the modeling of general relativistic polytropes in the presence of anisotropic pressure. As mentioned in the Introduction, we undertook this task, motivated on the one hand,  by the conspicuous presence of such an anisotropy in compact objects and its influence on their structure, and on the other by the fact that polytropes represent fluid systems with a wide range of applications in astrophysics (e.g. Fermi fluids).

Thus, we have identified two possible  polytropic equations of state. For each case we have found the full set of equations, which become the generalized Lane-Emden equation for the anisotropic matter in the Newtonian limit.

{As it should be expected (since an additional variable is introduced),  the above mentioned set of equations requires additional information to be integrated, . This information should be provided through the specific description of the anisotropy present in each problem. However in order to illustrate our formalism we have proceeded to adopt a specific ansatz in order to integrate the system for both cases. The motivation for such a choice was already explained, however it should be emphasized that the main reason  to present such models was not to describe any specific astrophysical scenario, but to convince the reader that the formalism  works.}

{Nevertheless the obtained models exhibit some  interesting features which deserve to be commented. We did so in the previous section with the help of the Tolman mass concept. Particularly interesting is the fact that some of the more relevant features of the models are present in models not related in any way to the ansatz used here.}

At this point, we envisage two possible lines of research  to apply the formalism presented in Sec. III, namely:
\begin{itemize}
\item to identify  the $\Delta$ associated  to a specific astrophysical problem from purely physics considerations and then proceed to integrate the corresponding set of  equations.
\item to introduce a heuristic ansatz (different from the one assumed here)  allowing one to integrate the system (\ref{TOV1h}), (\ref{veprimah}) or (\ref{TOV2h}), (\ref{veprima2h}). One example of which, could be the method proposed by Lake to obtain anisotropic solutions to Einstein equations from isotropic Newtonian solutions \cite{Lake}
\end{itemize}

Also, among the many unanswered  questions, related to the issue considered here, two are called particularly our attention, namely:
\begin{itemize}
\item is there a link between the two anomalous behaviours described in the previous section?
\item is the different response of the stability, to different degrees of anisotropy, somehow related to the possible ``cracking'' of the configurations as described in \cite{cr1, p3} (and references therein)?
\end{itemize}

Finally, we want to stress that all we have done here requires spherical symmetry, at least as an approximation. It is possible that this symmetry can be broken by a strong magnetic field, rendering the distribution anisotropic and nonspherical. In such a case, of course, the method presented here does not apply, or eventually applies only approximatively.

\section*{Acknowledgments}

{W. B. wishes to thank the Departamento de F\'\i sica Te\'orica e Historia de la Ciencia, Universidad del Pa\'\i s Vasco, for hospitality, especially J. 
Ib\'a\~nez and A. Di Prisco; and also the Intercambio Cient\'\i fico Program, U.L.A., for financial support.}

\end{document}